%% file: arXiv.tex
\documentclass[twocolumn]{aastex62}
\journalinfo{\@Draft}
\usepackage{natbib}
\usepackage{xcolor}
\usepackage{color}
\usepackage{comment}
\usepackage{amsmath}
 \usepackage{orcidlink}

\definecolor{royalfuchsia}{rgb}{0.79, 0.17, 0.57}
\definecolor{red}{rgb}{1.0, 0., 0.}
\renewcommand\tablename{Table}

\begin{document}

\title{IDENTIFICATION OF EXTENDED EMISSION GAMMA-RAY BURSTS CANDIDATES USING MACHINE LEARNING}

\author[0009-0001-2607-6359]{Garcia-Cifuentes,~K.\,\orcidlink{0009-0001-2607-6359}}
\affiliation{Instituto de Ciencias Nucleares,  Universidad Nacional Aut\'onoma de M\'exico, Apartado Postal 70-543, 04510 CDMX, M\'exico}

\author[0000-0002-0216-3415]{Becerra,~R.~L.\,\orcidlink{0000-0002-0216-3415}}
\affiliation{Instituto de Ciencias Nucleares, Universidad Nacional Aut\'onoma de M\'exico, Apartado Postal 70-264, 04510 M\'exico, CDMX, M\'exico;}

\author[0000-0002-3137-4633]{De~Colle,~F.\,\orcidlink{0000-0002-3137-4633}}
\affiliation{Instituto de Ciencias Nucleares, Universidad Nacional Aut\'onoma de M\'exico, Apartado Postal 70-264, 04510 M\'exico, CDMX, M\'exico;}

\author[0000-0001-6546-1180]{Cabrera,~J.~I..\,\orcidlink{0000-0001-6546-1180}}
\affiliation{Facultad de Ciencias, Universidad Nacional Aut\'onoma de M\'exico, A. P. 70-543, 04510 D.F, M\'exico}
\affiliation{Colegio de Ciencias y Humanidades Plantel Sur, Universidad Nacional Autónoma de México, 04500 Ciudad de México, México}

\author[0000-0002-8949-5200]{Del Burgo,~C..\,\orcidlink{0000-0002-8949-5200}}
\affiliation{Instituto Nacional de Astrof\'isica, \'Optica y Electr\'onica, Luis Enrique Erro 1, Sta. Ma. Tonantzintla, Puebla, M\'exico}

\begin{abstract}
Gamma-ray bursts (GRBs) have been traditionally classified based on their duration. The increasing number of extended emission (EE) GRBs, lasting typically more than 2 seconds but with properties similar to those of a short GRBs,  challenges the traditional classification criteria.
In this work, we use the t-Distributed Stochastic Neighbor Embedding (t-SNE), a machine learning technique, to classify GRBs.
We present the results for GRBs observed until July 2022 by the {\itshape Swift}/BAT instrument in all its energy bands. We show the effects of varying the learning rate and perplexity parameters as well as the benefit of pre-processing the data by a non-parametric noise reduction technique. 
Consistently with previous works, we show that the t-SNE method separates GRBs in two subgroups. We also show that EE GRBs reported by various authors under different criteria tend to cluster in a few regions of our t-SNE maps, and identify seven new EE GRB candidates by using the gamma-ray data provided by the automatic pipeline of {\itshape Swift}/BAT and the proximity with previously identified EE GRBs.
\end{abstract}

\begin{center}

\end{center}

\section{Introduction} \label{sec:intro}

Gamma-ray bursts (GRBs) are typically classified according to their duration $T_{90}$, i.e., the time over which a burst emits from 5\% to 95\% of its total observed counts \citep{kouveliotou1993}. 
Short GRBs (SGRBs), typically lasting $\lesssim 2$ s, are associated with the coalescence of compact objects, such as neutron stars or a neutron star and a black hole \citep{eichler89,narayan92,ruffert98,rosswog02,giacomazzo11,lee07}, while long GRBs (LGRBs), lasting $\gtrsim 2$ s, are associated to the collapse of stripped stars and their explosion as energetic type Ic supernovae \citep{woosley93,macfadyen99,hjorth12,becerra17}. The SGRBs gamma-ray emission is harder and dimmer, and they are typically seen in galaxies with low or absent star formation, while LGRBs have a softer spectrum, a larger total luminosity, and are typically seen in star-forming galaxies \citep{gehrels13}. That said, there is a certain overlap between the two classes, which makes it difficult to separate them, especially in GRBs lasting more than 2~s \citep{troja22,becerra23,becerra19a}.

The existence of a third population, the extended emission GRBs, has been proposed \citep{norris06}. This is characterized by the presence of a faint, soft X-ray and gamma-ray emission after the main peak in gamma-ray emission. Their identification is uncertain since their duration overlaps with that of LGRBs. However, their spectral properties 
 are more similar to those of short GRBs \citep{zhang06,becerra19b}. This apparent overlap challenges their classification, making them of great interest, due to the importance of restricting the progenitors that would give rise to this population.

\citet{jespersen20} demonstrated that a manifold learning algorithm can be employed to separate the two standard classes (short and long GRBs). Using the {\itshape Swift}/BAT GRB catalog, consisting of light curves in four energy bands (15-25~keV, 25-50~keV, 50-100~keV, 100-350~keV) for about 1250 events, they classified the events based on the ``t-Distributed Stochastic Neighbor Embedding'' (t-SNE) method \citep{maaten08}. Briefly, this is a dimensionality reduction method, which lessens the information from the initial N-dimensional space (where N is the number of data points in the light curves) to a two-dimensional (2d) space. 
The t-SNE method provides an easy way to find similarities between light curves of all the events from the sample, analyzing the close-by regions \citep{jespersen20,Dimple2023}.  

Unfortunately, the t-SNE algorithm leads to embeddings in a non-parametric form. Therefore, we cannot simply add new events to an existing study, as demonstrated in the cytometry analysis carried out by \cite{donnenberg07}. Furthermore, although the clustering of two subgroups has been observed in \citet{jespersen20}, little attention has been paid to the embedding details (such as t-SNE optimization parameters), resulting in poor visualization map repeatability.

Here, we present visualization maps and an in-depth analysis of t-SNE embeddings on the entire available {\itshape Swift} catalog (updated to July 2022). Our approach is focused on giving a detailed analysis of t-SNE optimization and on showing that adding a non-parametric noise reduction technique ({\sc FABADA};  \citealt{sanchez22}) in the pre-processing step improves the resulting visualization maps without changing their overall structure. Additionally, we analyze the extended emission sub-samples identified previously through a diversity of criteria. The aim of this work is to investigate if such events can be recognized in any way from their gamma-ray emission using only the {\itshape Swift}/BAT data. In addition, throughout this work, we seek to identify candidates for GRBs with extended emission using t-SNE visualization maps.

This paper is organized as follows. In \S\ref{sec:methodology}, we present our methodology. In \S\ref{sec:resclassification}, we analyze previous classifications of GRBs. In \S\ref{sec:resee}, we discuss the criteria used to recognize GRBs with extended emission, and we identify candidates for this group using the proximity criterion with previously identified clusters of objects when visualizing them in t-SNE maps. Finally, in \S\ref{sec:summary}, we summarize our results.

\section{Methodology} 
\label{sec:methodology}

We use the four light curves from the {\itshape Swift}/BAT GRB Catalog\footnote{\label{footnote1}\url{https://swift.gsfc.nasa.gov/results/batgrbcat/}} (15-25 keV, 25-50 keV, and 50-100 keV, and 100-350 keV).
 In total, we use 1527 light curves in four energy bands and in their sum. To compare our results with those of \citet{jespersen20}, we use the pre-processing method outlined in their study.

\begin{enumerate}
    \item The light curves are filtered using the $T_{100}$ duration presented in the {\itshape Swift}/BAT GRB Catalog. We removed 74 GRBs from the 64~ms dataset for the following reasons: 27 GRBs have less than 3 points per light curve, and 47 GRBs do not have a reported duration $T_{100}$ in the {\itshape Swift}/BAT summary table or lack event data. 
    \item We normalized the light curves by the total fluence, which was calculated by integrating over the $15-350$~keV band using the ``Simpson rule" implemented in the Python3 library {\sc scipy.integrate}.
    \item We ensured that every GRB has the same number of data points by padding zeros after the trigger in all bands.
    
    \item We perform discrete-time Fourier transform (DTFT) on all events using {\sc scipy.fft}, concatenating the data of all bands in ascending order of energy for each GRB.
    
    \item We applied the t-SNE method to the Fourier Amplitude Spectrum for the entire sample. We follow the same procedure implemented in the t-SNE instance of \textit{scikit Learn} \citep{pedregosa12}: We set initial coordinates from an isotropic Gaussian random initialization. Then, we start the embedding on an early exaggeration phase ($EE$) of $250$ steps with an early exaggeration factor ($ee$) of $12$. Finally, we finish the embedding with a regular phase of $750$ steps disabling the early exaggeration factor and using a perplexity of $5$ or $30$ (in both phases, we set a fixed learning rate $\eta$ of $200$). In addition, we set {\sc method}$=$\textit{exact} to get the most accurate results, and for repeatability, we set {\sc random\_state}$ = 42$.
\end{enumerate}

In addition to using the 64~ms dataset, we also employ the $S/N>5$/10~s binning data (hereafter, we will refer to this data set as 10-s binning data set), which are not equally spaced (we removed 199 GRBs from the 10~s dataset for the reasons pointed out before). Therefore, we perform linear interpolation at a fixed resolution of $13$ ms to ensure a common axis  (this parameter was fixed by the computational resources and memory available).
Additionally, we perform a noise reduction algorithm based on Bayesian inference to improve the signal-to-noise ratio (S/N) of $64$~ms light curves. The procedure and results are described in detail in \S~\ref{sec:noise} and in the url of the \ {\sc ClassiPyGRB} package \footnote{\label{footnote2}\url{https://github.com/KenethGarcia/ClassiPyGRB}} (Garcia-Cifuentes et al. 2023b in prep.)
 
\subsection{t-SNE optimization}
\label{sec:t-SNE}

The methodology presented in \S~\ref{sec:methodology} produces 2D visualization maps colored by the duration $T_{90}$. We stand out that in non-linear dimensional reduction techniques, such as t-SNE, ``x'' and ``y'' axes do not have any significance or units; only the structure on the visualization map is meaningful. Therefore, getting the same shape rotated in different figures does not have any physical interpretation.

We note that the light curves pre-processed by the {\itshape Swift} pipeline should only be considered for energies in the interval between 15 and 150~keV. This follows from three considerations: i) although there is a wider energy range, the three first channels are below that of the FHMW, ii) the sensitivity of these channels is limited by the threshold determined by the electronics above 150~keV \citep{krimm04}, the encoded mask becomes transparent to photons, as well as being below the FHMW, and iii) the calibration for energies above 150~keV is no longer reliable since the radioactive calibration source for {\itshape Swift}/BAT is a source of $^{241}$Am whose main lines are 26.3~keV and 59.5~keV. Nevertheless, we decided to use the data from the four bands in this work following \citet{jespersen20}, including the 100-350 keV energy band which partially falls outside the 15-150 keV energy limits.

\subsubsection{Perplexity role}

The main parameter to be defined when applying the t-SNE method is the \emph{perplexity} ($pp$), which determines the number of nearest neighbors considered for each GRB in the embedding \citep{maaten08}. For example, on the 1527 GRBs considered in this work, $16$ nearest neighbors are considered for each GRB when $pp$ is $5$ and $91$ for $pp=30$. 

Although the visualization map for $pp=30$ shows two clear subgroups in \citet{jespersen20}, we noted a difference when performing t-SNE on the new data set considered in this paper: the evident absence of two subgroups for $pp=30$ embedding. We illustrate this in Figure~\ref{fig:1}. We observe that for $pp=5$ (panel a), there are two subgroups separated. Nevertheless, for $pp=30$ (panel b), the map shows a unique condensed cluster. We hypothesize that this situation may be due to two reasons. Firstly, a problem in the pre-processing procedure because using DTFT could not be the best pre-processing solution (as mentioned by \citealt{jespersen20}). Secondly, because $64$~ms data is quite noisy, the embedding can be distracted by joining two or more different subgroups in a single cluster.

Despite having a difference in the $pp=30$ embedding, we replicated the results reported by \citet{jespersen20}: i) the duration $T_{90}$ and the position of each GRB in the t-SNE maps are highly correlated; ii) t-SNE can distinguish two subgroups of GRBs (``short'' and ``long'') characterized by having a different distribution of duration, even though we extend the dataset from {\itshape Swift}/BAT up to July 2022. Therefore, t-SNE allows us to differentiate two kinds of GRBs without the need of controlled supervision in the algorithm or any prior assumption in the number of subgroups to find.   

\subsubsection{Noise Reduction Approach}
\label{sec:noise}

As a consequence of the few photons received in the gamma frequencies, several GRB light signals are dominated by noise. To handle this bias, we use two approaches: Firstly, we applied the non-parametric noise reduction technique implemented by {\sc FABADA} \citep{sanchez22} to each band for every single light curve (Figure \ref{fig:GRB060614}). Secondly, we use the 10~s binned light curve data from the {\itshape Swift}/BAT catalog.

Briefly, the {\sc FABADA} algorithm uses a Bayesian inference approach to handle noise, obtaining a statistically compatible estimation of the underlying signal. It only requires as arguments the data/image itself (i.e., two or three-dimensional data) and its computed noise variance. The noise variance is calculated using the root-mean-square (RMS) noise outside $T_{100}$ following \citet{immerkaer96}. 

On the other hand, we use the 10~s binning light curve data from the {\itshape Swift}/BAT catalog. Nevertheless, the 10~s binning data has the drawback of reducing the sample size from $1453$ to $1328$ GRBs (affecting mainly the sample of GRBs with $T_{90}<2~s$).

\begin{figure}
    \centering
    \includegraphics[width=\linewidth]{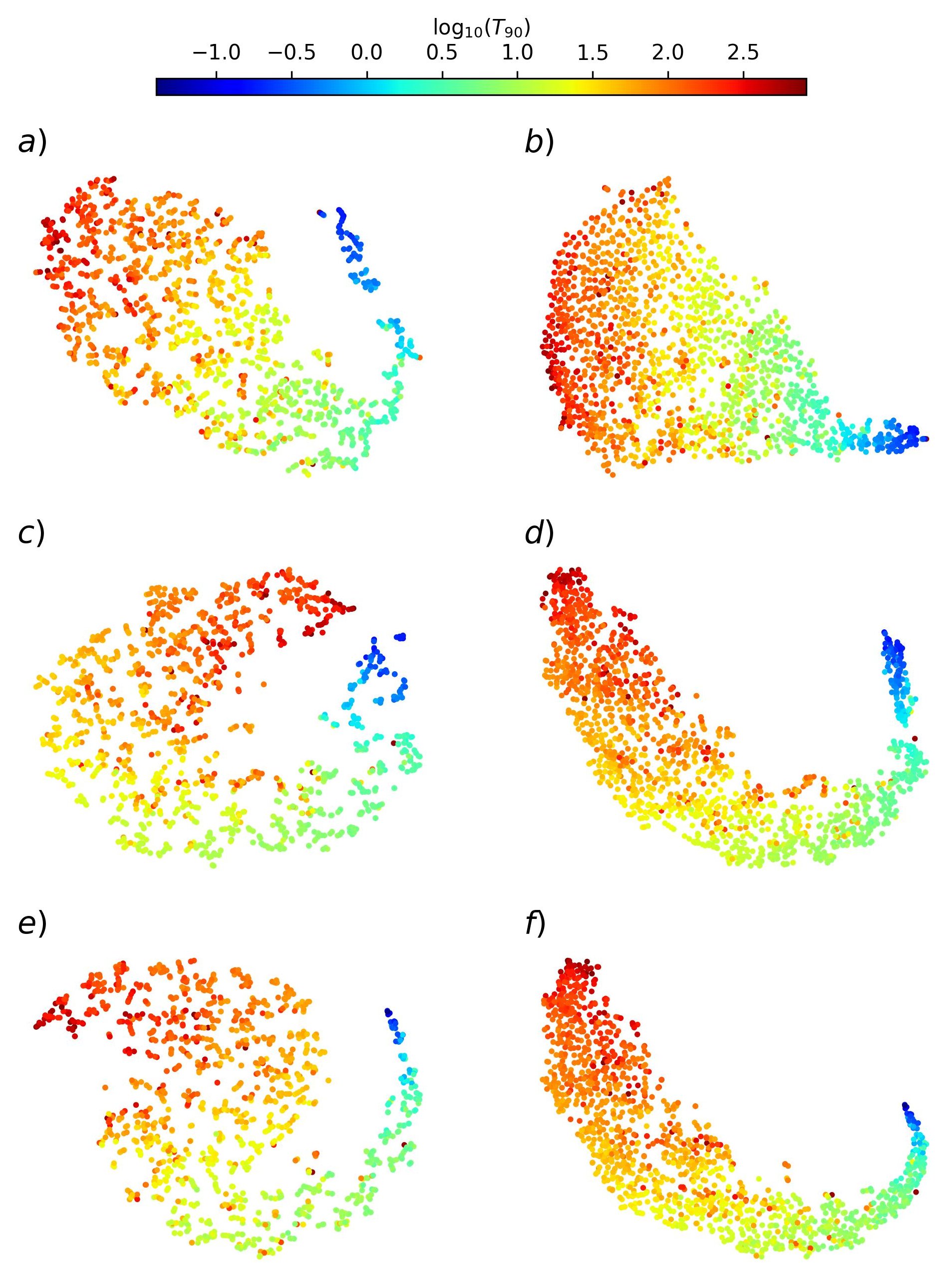}
    \caption{t-SNE visualization maps of processed Swift/BAT light curves using reference optimization values from \S~\ref{sec:methodology}. Dataset binned at 64~ms with \textbf{a)} $pp=5$ and \textbf{b)} $pp=30$. Noise-reduced dataset binned using {\sc FABADA} at 64~ms with \textbf{c)} $pp=5$ and \textbf{d)} $pp=30$. Dataset with S/N ratio higher than 5~s or 10~s binning with \textbf{e)} $pp=5$ and \textbf{f)} $pp=30$. The color bar shows the duration $T_{90}$ of the GRBs.}
    \label{fig:1}
\end{figure}

\begin{figure}
    \centering
    \includegraphics[width=\linewidth]{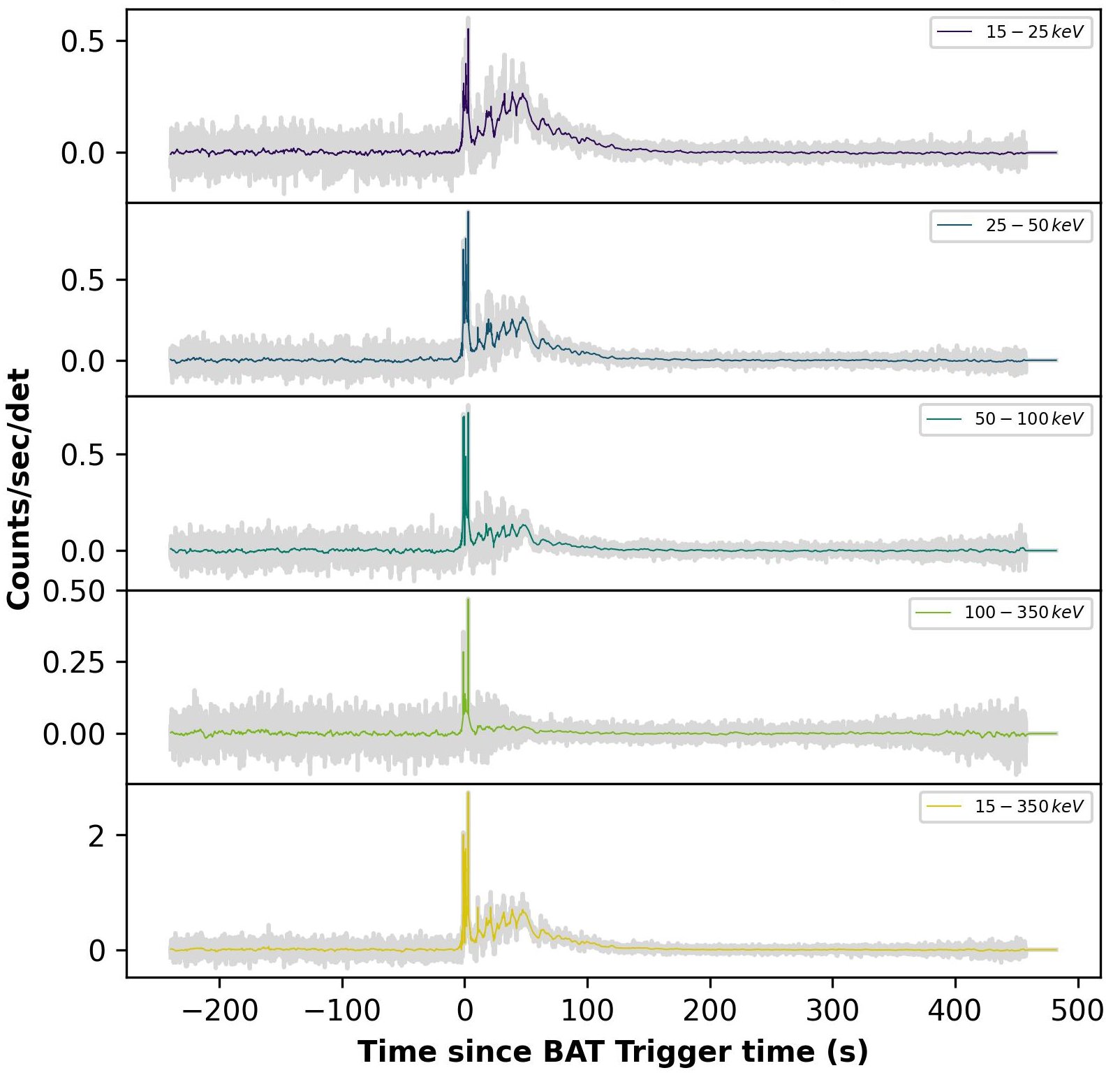}
    \caption{GRB 060614 light curves from {\itshape Swift} (gray line) and its noise-reduced counterpart obtained (colored lines) with the {\sc FABADA} \citep{sanchez22} algorithm.}
    \label{fig:GRB060614}
\end{figure}

After getting noise-reduced light curves (either by using {\sc FABADA} or $10$~s binning), we perform t-SNE applying the process described in \S~\ref{sec:methodology} to these light curves (i.e., see Figure~\ref{fig:GRB060614}). In panels c-f of Figure~\ref{fig:1}, we compare the different cases. The embeddings obtained for the {\sc FABADA} noise-reduced cases (panels c and d) show the most evident separation between the two main subgroups. Even for $pp=30$, these subgroups are discernible, improving the overall t-SNE maps, and therefore providing a more reliable classification. 

The use of noise-reduced datasets has some trade-offs depending on the underlying nature of the reducing process. This is illustrated in panels e and f of Figure~\ref{fig:1},  presenting the embeddings obtained by considering the 10~s binned dataset. Although these embeddings have the same general trend (e.g., with respect to the duration $T_{90}$), we do not observe any evident subgroups/clusters present. 

We suggest that this difference is due to the decrease in the $T_{90} < 2~s$ GRB sample due to the 10~s binning: we went from having 106 (with 64~ms samples) to 38 (only $\sim 36\%$ of the 64~ms sample). By contrast, the GRB sample having $T_{90} > 2$~s went from having $1346$ at 64~ms to 1290 at 10~s (a decrease of only $\sim 4\%$). Of these significant changes in the sample, the SGRBs cluster is not well discerned by t-SNE. Then, the embeddings are worse than in the noise-reduced case using {\sc FABADA}.

\section{GRB classification}
\label{sec:resclassification}

\begin{figure*}
    \centering
    \includegraphics[width=1\linewidth]{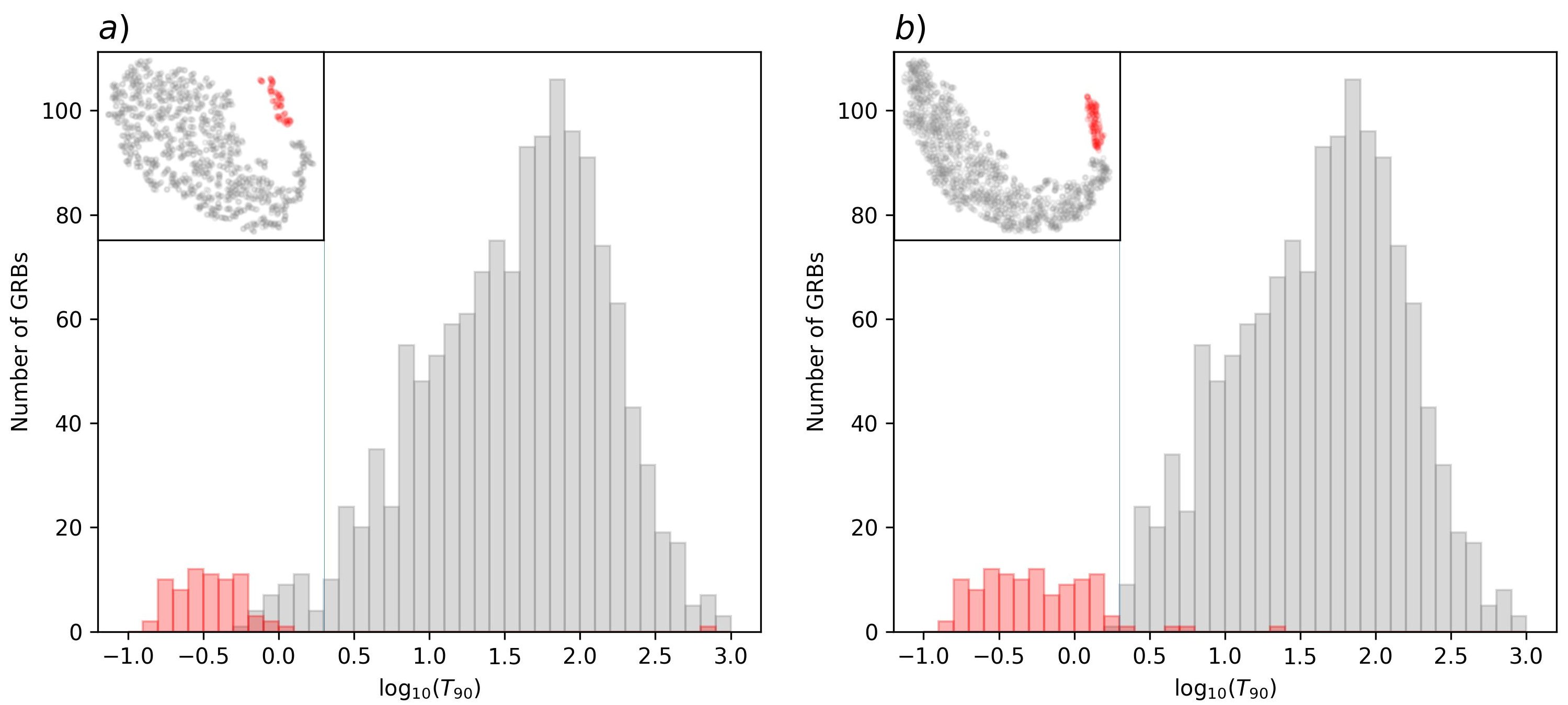}
    \caption{\textbf{a)} Distribution of durations ($T_{90}$) for clusters discerned by t-SNE on the 64~ms dataset. \textbf{b)} Same as panel a), but using the {\sc FABADA} noise-reduced dataset ($pp=30$). Upper-left insets shows the result of the t-SNE method in the two cases. In all figures, LGRBs/SGRBs are shown in gray/red. The gray vertical line represents the cut-off at $T_{90} = 2$~s frequently used in literature to distinguish between long and short GRBs. }
    \label{fig:histograms}
\end{figure*}

We refine the t-SNE embeddings by using {\sc FABADA} to increase the S/N. As shown in figure \ref{fig:1}, panels c and d, GRBs are clustered in two sub-groups. To understand the physical meaning of these subgroups we need to analyze their physical properties.
The effect of using a noise-reduction technique to pre-process the data is illustrated in Figure~\ref{fig:histograms}. This figure presents the distribution of GRB durations in the 64~ms dataset with and without using {\sc FABADA}. In the figure, SGRBs are represented in red and LGRBs in gray.
The figure considers only the $pp=5$ case for 64~ms data and $pp=30$ for {\sc FABADA} reduced dataset (where the 2D plots resulting from the t-SNE show two well-defined groups in both cases). 
The existence of a third group formed by the EE GRBs and their possible clustering will be discussed in \S~\ref{sec:resee}.

Figure \ref{fig:histograms} shows that classifications based on model-independent data analysis as t-SNE are strongly influenced by the presence of noise. Performing t-SNE in the {\sc FABADA} noise-reduced dataset changes the classification of 40 GRBs: 39 bursts switch from the long to the short group (most of them have a duration between $\sim$ 1 and $\sim$ 2~s), while only one burst switch from short to long (GRB 190718A). The effective population of short GRBs increases 53\%, a significant loss that reveals the importance of reducing noise when machine learning approaches are used to classify GRBs.

\input{table_classification}

The 64~ms and FABADA noise-reduction methods catalog 25 and 11 events differently in comparison to \citet{steinhardt2023}.
The full list of these events is shown in Table~\ref{table:classification}. In the classification based on the {\sc FABADA} noise-reduced sample (see Figure~\ref{fig:1}, panels d and e) 6/5 long/short GRBs were classified as short/long by \citet{steinhardt2023}; whereas, in the classification based on 64~ms data, 25 long GRBs were classified as short by \citet{steinhardt2023}.

Moreover, we observe in Figure~\ref{fig:histograms} the presence of long GRBs with $T_{90}$ between 4 and 800~s in the ``short'' cluster categorized by t-SNE. GRB 190718A ($T_{90}=704$~s) is the singular case in the 64~ms data (panel a), while for {\sc FABADA} dataset are GRB 171103A ($T_{90}=4.27$~s), GRB140209A ($T_{90}=20.432$~s), and GRB090510 ($T_{90}=5.66$~s, with EE).

\input{eeothers.tex}

\section{GRBs with extended emission}
\label{sec:resee}

Several authors have reported the existence of a third class of GRBs. These GRBs present an extended emission (EE) in gamma rays, although at lower energies in comparison with other LGRBs. Most of them can be distinguished by long and short GRBs because they show a bright episode \citep{norris96} after the initial prompt emission and an intermediate quiescent phase. One of the most recent example of EE GRB is GRB 211211A. This event has $T_{90}=13$~s, an extended emission duration of $\sim$ 55~s \citep{Yang2022}, and it has been associated with kilonova emission \citep{troja22}.
GRB 211211A presented a harder emission than GRB 060614 in the EE phase, which the closest analogy. GRB 060614, whose duration and extended emission duration are is $T_{90}=6$~s and 100~s respectively, was classified as a long-duration Type-I burst whose temporal lag and peak luminosity fall entirely within the short-duration GRB subclass \citep{Gehrels2006}. A detailed comparison between these two events is presented in Table~1 of \cite{Yang2022}.

Previous works dedicated to the classification of EE GRBs used a broad range of classification criteria, demonstrating that correctly identifying the elements of this subclass is challenging.

\cite{norris10} suggested a Bayesian block treatment to discern EE in {\itshape Swift}/BAT GRBs, classifying 12 out of 51 GRBs with EE under this criterion. On the other hand, \cite{kaneko15} considered GRBs with a main burst lasting $\lesssim 5$ s followed by fainter emission identified by considering a S/N$>1.5\;\sigma$ in 4~s binned light curves, being $\sigma$ the background fluctuation, and finding 16 {\itshape Swift}/BAT and {\itshape Fermi}/GBM EE GRBs. In the data release of {\itshape Swift}/BAT, \cite{lien16} presented a list of ``definite'' EE GRBs based on a qualitative criterion. They double-checked each GRB by eye inspection based on information from GCN circular reports. \cite{kisaka17} used a phenomenological formula to fit the EE in {\itshape Swift}/BAT and {\itshape Swift}/XRT data and found that about half of the total GRB sample had an EE. It is important to notice that their model classified EE GRBs based mainly on X-ray emission detected by XRT. XRT data are not included in the analysis presented in this paper.

\cite{dichiara21} showed that {\itshape Swift}/BAT detectors could not take into account signal from EE if the source is observed during slewing intervals. To compensate for this effect, they collected slew-corrected images from 2~s to 150~s after the trigger time for a sample of 8 high redshift GRBs, classifying 5 of them as EE GRBs. Furthermore, they showed the importance of low S/N light curves in GRB classification and their EE detection. In fact, in GRBs localized at large distances the EE can be lower than the background noise, being undetectable.

\begin{figure*}
    \centering
    \includegraphics[width=1\linewidth]{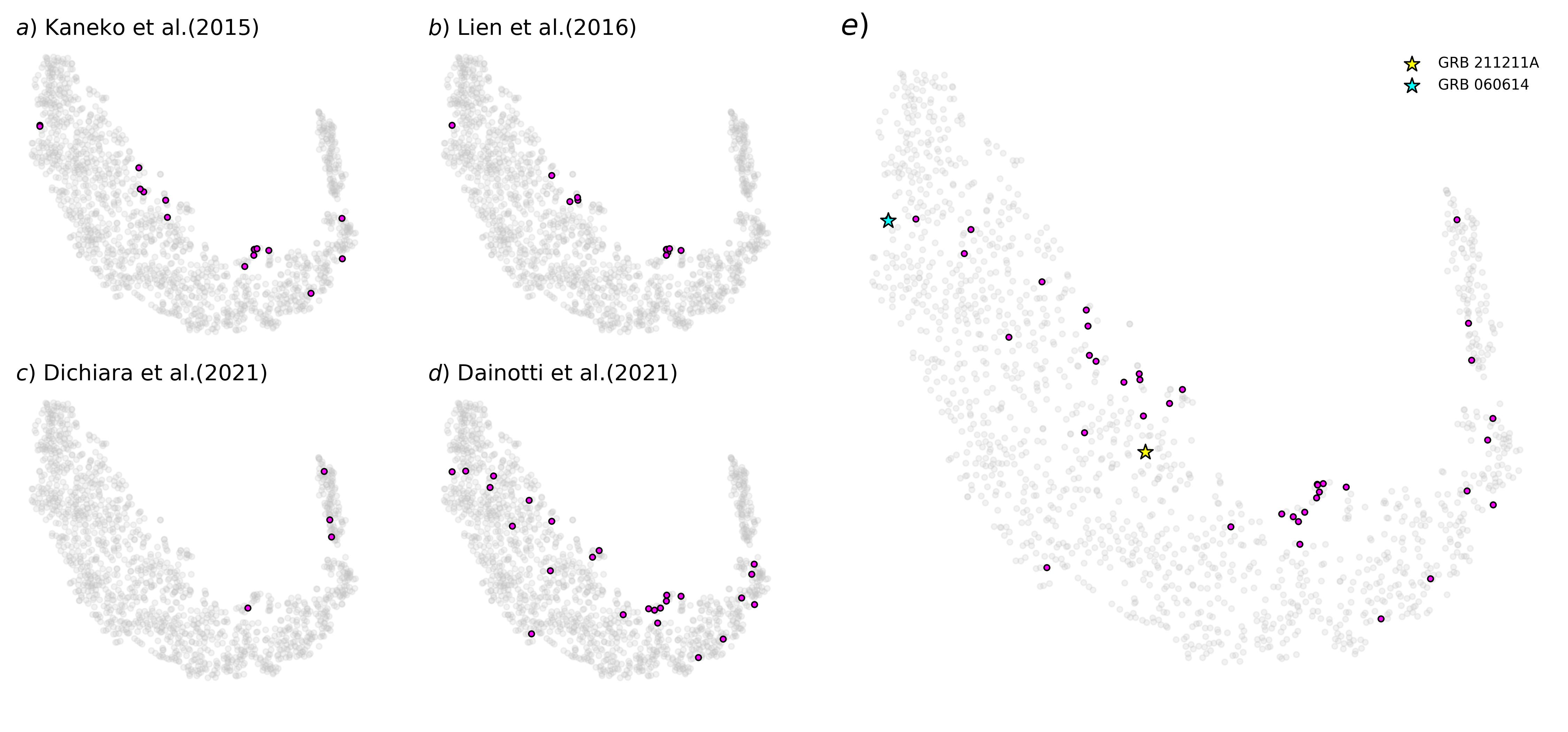}
    \caption{t-SNE visualization map obtained for the noise-reduced dataset binned at $64\,ms$ with $pp=30$. GRBs colored in magenta are classified as Extended Emission by \textbf{a)} \cite{kaneko15}; \textbf{b)} \cite{lien16}; \textbf{c)} \cite{dichiara21}; \textbf{d)} \cite{dainotti21}; \textbf{e)} All EE reported from panel a-d. We highlight the position of GRB 060614 (cyan star) and GRB 211211A (yellow star).}
    
    \label{fig:others}
\end{figure*}

Other works also used qualitative criteria to identify EE, such as the extensive search on the Swift database performed by \cite{dainotti21}. They used the sample of EE reported by \cite{norris96} with known redshifts and information about the peak flux and the spectral features in order to have a more “complete” sample. On the other hand, \cite{zhang20} applied a S/N criterion (similarly to \citealt{kaneko15}) to previously reported EE GRBs. We show all the EE identified in these previous works in Tables~\ref{tab:eelist} and~\ref{tab:eelist1}, and we summarize the criteria used for each group in Table~\ref{tab:criteria}. Figure~\ref{fig:others} illustrates GRBs with EE analyzed previously as highlighted points in t-SNE maps in each case.

\input{tableee2.tex}

Focusing on studies based only in inspection/analysis of {\itshape Swift}/BAT light curves (panels a, b, and d of Figure~\ref{fig:others}), EE GRBs are distributed almost exclusively over the long GRB subgroup in the t-SNE maps. In addition, they are highly clustered along a few regions, independent of the considered case. For instance, in the case of the EE GRBs classified by \cite{lien16} and shown in panel b of figure~\ref{fig:others}, 6 EE GRBs are clustered at the bottom right of the LGRB subgroup (5 of which GRBs are nearly overposed in the figure), 3 are clustered in the middle of the plot, and two remain isolated at the top of the figure. A similar clustering (although with a larger dispersion) is present in the EE GRBs shown in panel a, c and d. In all cases, a large number of EE GRBs remain clustered in the same regions as those of panel b).
This behavior is an expected property of the t-SNE algorithm: bursts sharing properties of both short and long GRBs (as in EE GRBs) are placed in the edges of the most similar cluster (in almost all cases, the long GRB subgroup).

\subsection{Identification of EE GRB candidates}
\label{sec:candidates}

Based on the properties of EE GRBs and their apparent clustering by t-SNE, it is possible to classify and identify candidates of EE GRBs by looking at the nearest neighbors of previous EE candidates reported. We carried out a search of neighbors of previous EE GRBs in panels a, b, c and d of Figure~\ref{fig:others}).  These GRBs were identified by a qualitative analysis of the light curves in the 4 available bands, and by a comparison with previously classified EE GRBs. We list in Table~\ref{tab:candidates} the seven candidates to GRBs with EE.

Within this sample, GRB 080123 has not been identified as an EE in previous papers. Nevertheless, it was pointed out by \citet{ibrahim08} as a potential event from an emerging class of short bursts with EE and as a possible GRB with EE.
Moreover, the extended emission of GRB 180618A \citep{Jordana-Mitjans2022,Fong2022,oconnor2022}, GRB 170728B \citep{Fong2022,oconnor2022}, GRB 180805B \citep{Fong2022,oconnor2022} and GRB 200219A was suggested in the GCN/TAN circulars.



\input{candidates.tex}

Finally, we notice that the classification is uncertain. An example is the GRB 090530, identified previously as a GRB with EE by \cite{dainotti21} and \cite{kaneko15} and located in the bottom right cluster of \citet{lien16} EE GRBs. Nevertheless, we did not find in the 64~ms or 10~s light curves any signature to support this classification as well as no evidence of the events surrounding it. We suggest that this is because of the identification of GRBs with EE using information from {\itshape Swift}/XRT, which datasets are not included in the analysis presented in this paper. 
This is a clear example of the disadvantages of not having a homogeneous criterion.

\section{Summary and conclusions} \label{sec:summary}

\citet{jespersen20} proposed the use of the t-SNE method to classify GRBs based on gamma-ray light curves. 
In this study, we extend their analysis by employing light curves observed by {\itshape Swift}/BAT until July 2022, and visualizing them through the same t-SNE dimensionality reduction algorithm. 

We showed that non-linear dimensionality algorithms such as t-SNE are sensitive to
the implementation of the method and to the choice of the parameters used. In particular, the two subgroups (corresponding to the short and long classes of GRBs) were present for a perplexity of $pp=30$ in \citet{jespersen20}, and $pp=5$ in our case.
We also found that by introducing a noise-reduction algorithm the embeddings obtained show two well-defined groups independent of the perplexity used.

As shown in Figure~\ref{fig:histograms}), some GRBs with a duration $T_{90}>4$~s are classified by the t-SNE method as part of the short GRBs subgroup (consistently with \citealt{steinhardt2023}).
The distribution of short GRBs with long duration is not continuous (whereas the same distribution is continuous for {\itshape BATSE} and {\itshape Fermi}, see \citealt{steinhardt2023}, Figure~7). 
This is related to the energy frequencies of the {\itshape Swift}/BAT and therefore, to the lower probability to detect such events in comparison with {\itshape BATSE} and {\itshape Fermi} (which detectors are optimized to higher frequencies). 
Thus, if the sample number of short GRBs detected by {\itshape Swift}/BAT were to increase, a more continuous Gaussian distribution would be seen.

We also presented an analysis of bursts previously classified as extended emission GRBs.
Although different criteria have been used in the literature for their classification, tend to cluster in very well-defined regions, located at the edges of the two-dimensional distribution of LGRBs generated by t-SNE.
This shows the similar photometry of these types of events. 

Although extended emission GRBs do not form a separated cluster, it is interesting that they cluster in several sub-groups (see, e.g., panel b of figure \ref{fig:others}
 where most EE GRBs are clustered in two regions of the map). By looking at the proximity of GRBs with previously identified clusters,
we have been able to identify seven new GRB candidates with EE. This method leads to reducing the time invested in the eye inspection for their classification and without the bias introduced by the human factor. 

\section*{ACKNOWLEDGEMENTS}
The authors thank the anonymous referee for the helpful comments that improved the quality of the manuscript.
We acknowledge the use of public data from the Swift data archive. 
We acknowledge support from the UNAM-PAPIIT grant AG100820. KSGC acknowledges support from CONACyT fellowship. RLB acknowledges support from CONACyT postdoctoral fellowship. CdB acknowledges support by Mexican CONACYT research grant FOP16-2021-01-320608.

We acknowledge the useful comments and suggestions provided by Antonio Castellanos-Ram\'irez.



\newpage
\renewcommand{\thesection}{\Alph{section}}
\setcounter{section}{0}

\section{Light curves of candidates to extended emission GRBs}

To show the observational evidence of an extended emission signature of the seven candidates identified in this work (see Table~\ref{tab:candidates}), we present their {\itshape Swift}/BAT light curves.

 \begin{figure}[b]
     \centering
     \includegraphics[width=\linewidth]{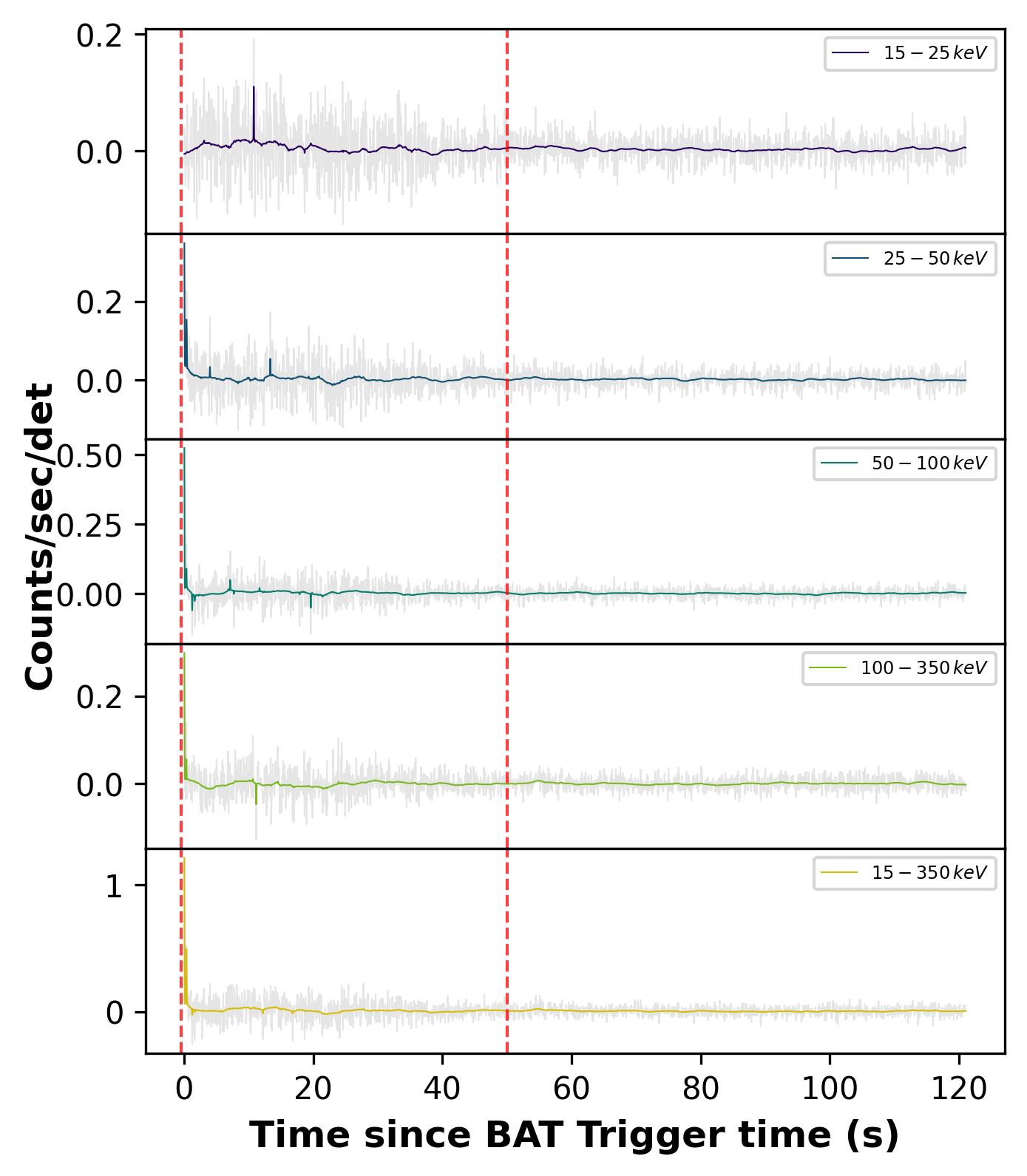}
     \caption{Light curves of GRB 080123 as extended emission GRB identified in this work. The light curves are taken from {\itshape Swift}/BAT (gray line) and its noise-reduced counterpart  (colored lines) were obtained with the {\sc FABADA} \citep{sanchez22} algorithm, limited to $T_{100}$. The extended emission region identified is illustrated with dashed red lines, with the lower temporal limit shifted $0.5$~seconds for better visualization of the main peak.}
     \label{fig:GRB 080123}
 \end{figure}
 
 \begin{figure}[b]
     \centering
     \includegraphics[width=\linewidth]{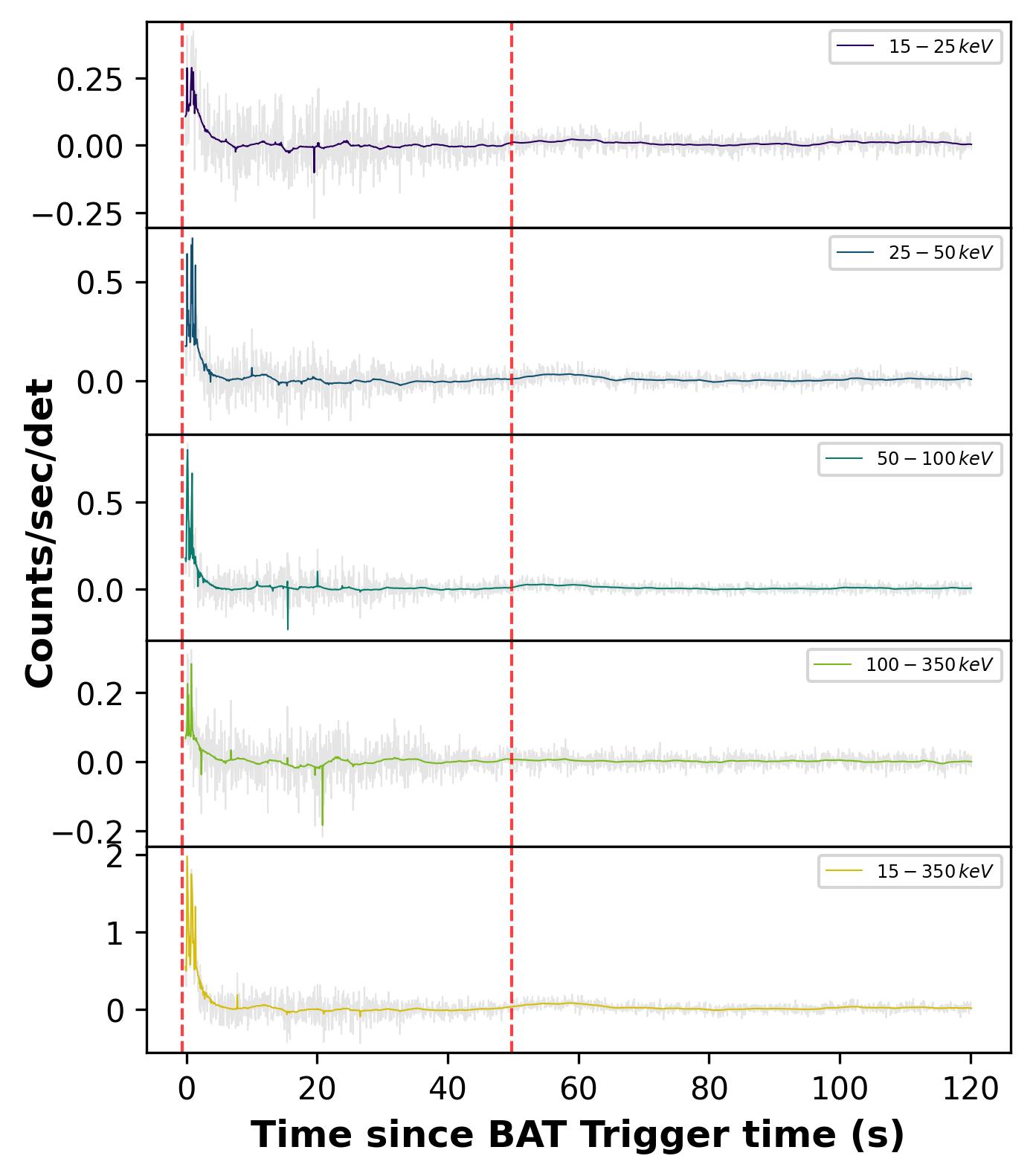}
     \caption{Light curves of GRB 140506A as extended emission GRB identified in this work. The light curves are taken from {\itshape Swift}/BAT (gray line) and its noise-reduced counterpart  (colored lines) were obtained with the {\sc FABADA} \citep{sanchez22} algorithm, limited to $T_{100}$. The extended emission region identified is illustrated with dashed red lines, with the lower temporal limit shifted $0.5$~seconds for better visualization of the main peak.}
     \label{fig:GRB 140506A}
 \end{figure}
 
 \begin{figure}[b]
     \centering
     \includegraphics[width=\linewidth]{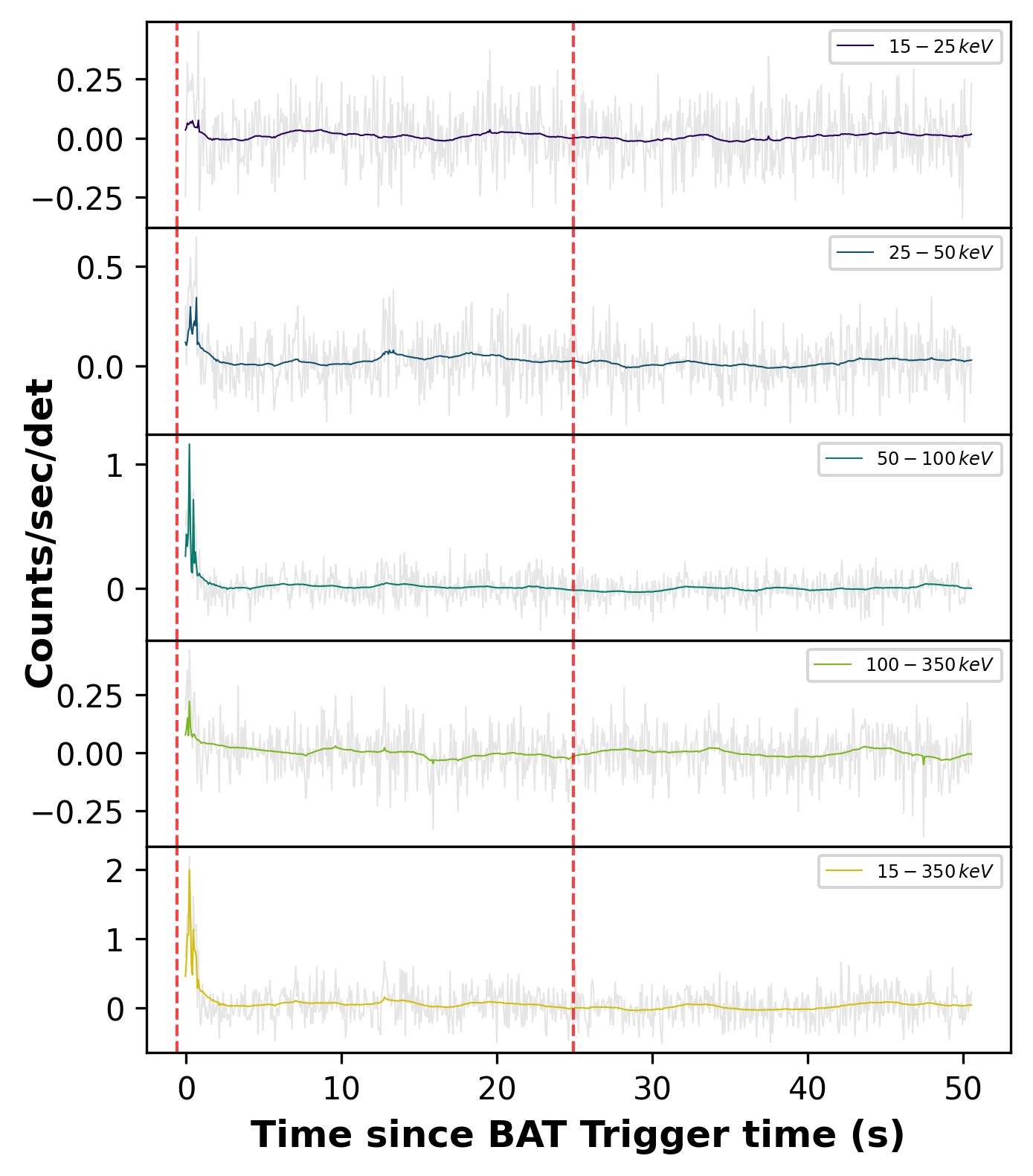}
     \caption{Light curves of GRB 170728B as extended emission GRB identified in this work. The light curves are taken from {\itshape Swift}/BAT (gray line) and its noise-reduced counterpart  (colored lines) were obtained with the {\sc FABADA} \citep{sanchez22} algorithm, limited to $T_{100}$. The extended emission region identified is illustrated with dashed red lines, with the lower temporal limit shifted $0.5$~seconds for better visualization of the main peak.}
     \label{fig:GRB 170728B}
 \end{figure}
 
 \begin{figure}[b]
     \centering
     \includegraphics[width=\linewidth]{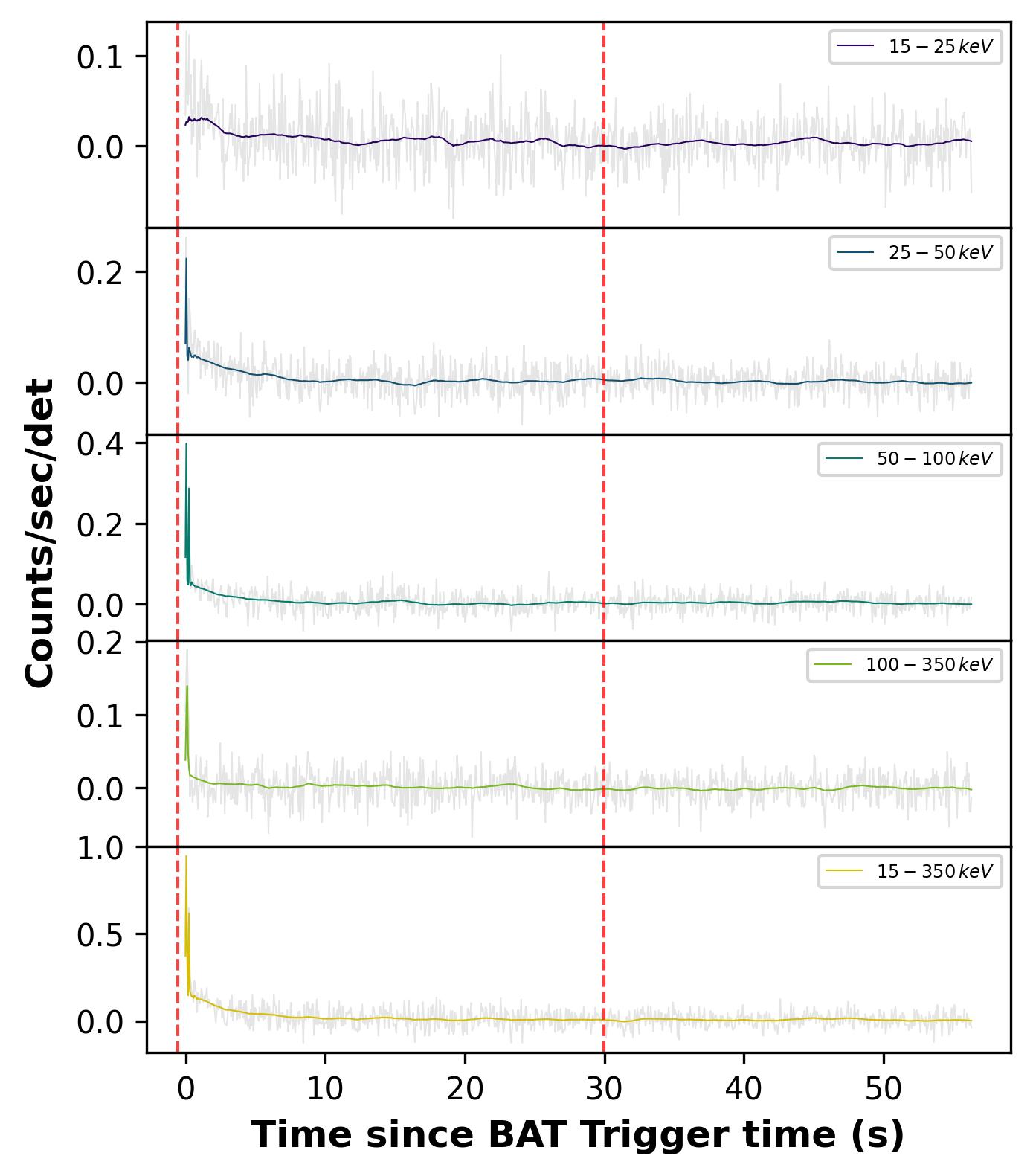}
     \caption{Light curves of GRB 180618A as extended emission GRB identified in this work. The light curves are taken from {\itshape Swift}/BAT (gray line) and its noise-reduced counterpart  (colored lines) were obtained with the {\sc FABADA} \citep{sanchez22} algorithm, limited to $T_{100}$. The extended emission region identified is illustrated with dashed red lines, with the lower temporal limit shifted $0.5$~seconds for better visualization of the main peak.}
     \label{fig:GRB 180618A}
 \end{figure}

\begin{figure}[b]
     \centering
     \includegraphics[width=\linewidth]{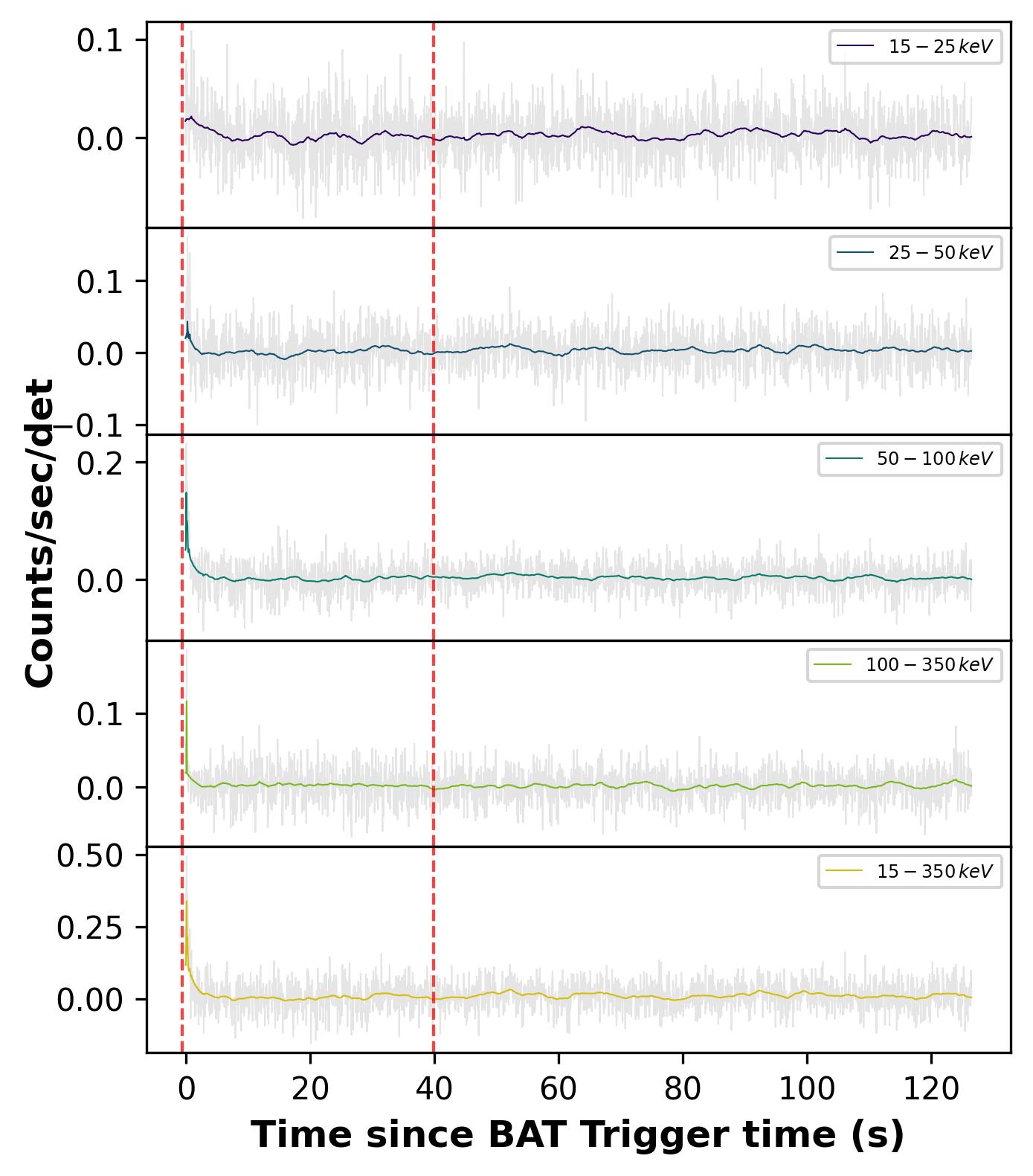}
     \caption{Light curves of GRB 180805B as extended emission GRB identified in this work. The light curves are taken from {\itshape Swift}/BAT (gray line) and its noise-reduced counterpart  (colored lines) were obtained with the {\sc FABADA} \citep{sanchez22} algorithm, limited to $T_{100}$. The extended emission region identified is illustrated with dashed red lines, with the lower temporal limit shifted $0.5$~seconds for better visualization of the main peak.}
     \label{fig:GRB 180805B}
 \end{figure}
 
 \begin{figure}[b]
     \centering
     \includegraphics[width=\linewidth]{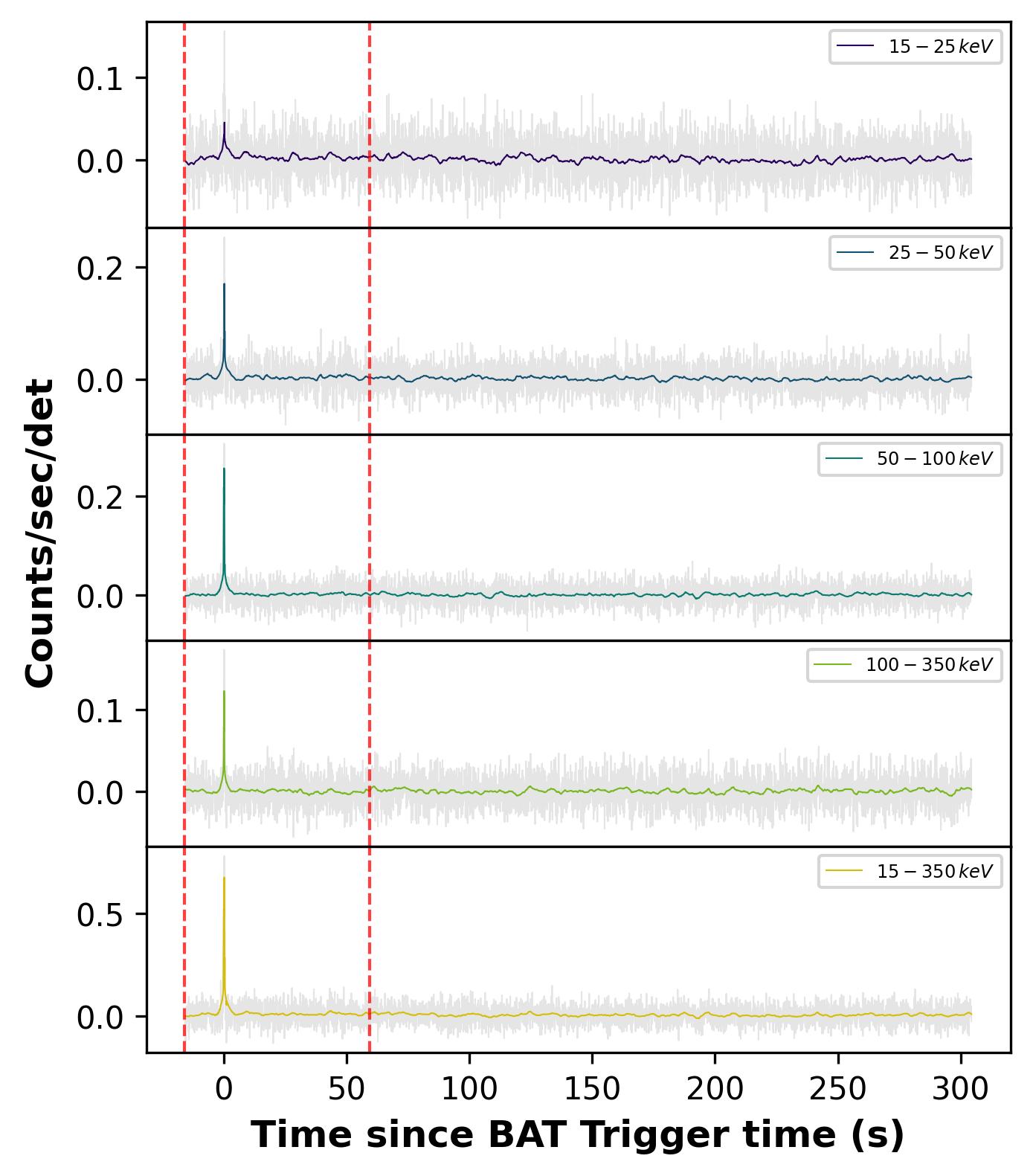}
     \caption{Light curves of GRB 200219A as extended emission GRB identified in this work. The light curves are taken from {\itshape Swift}/BAT (gray line) and its noise-reduced counterpart  (colored lines) were obtained with the {\sc FABADA} \citep{sanchez22} algorithm, limited to $T_{100}$. The extended emission region identified is illustrated with dashed red lines, with the lower temporal limit shifted $0.5$~seconds for better visualization of the main peak.}
     \label{fig:GRB 200219A}
 \end{figure}
 
 \begin{figure}[b]
     \centering
     \includegraphics[width=\linewidth]{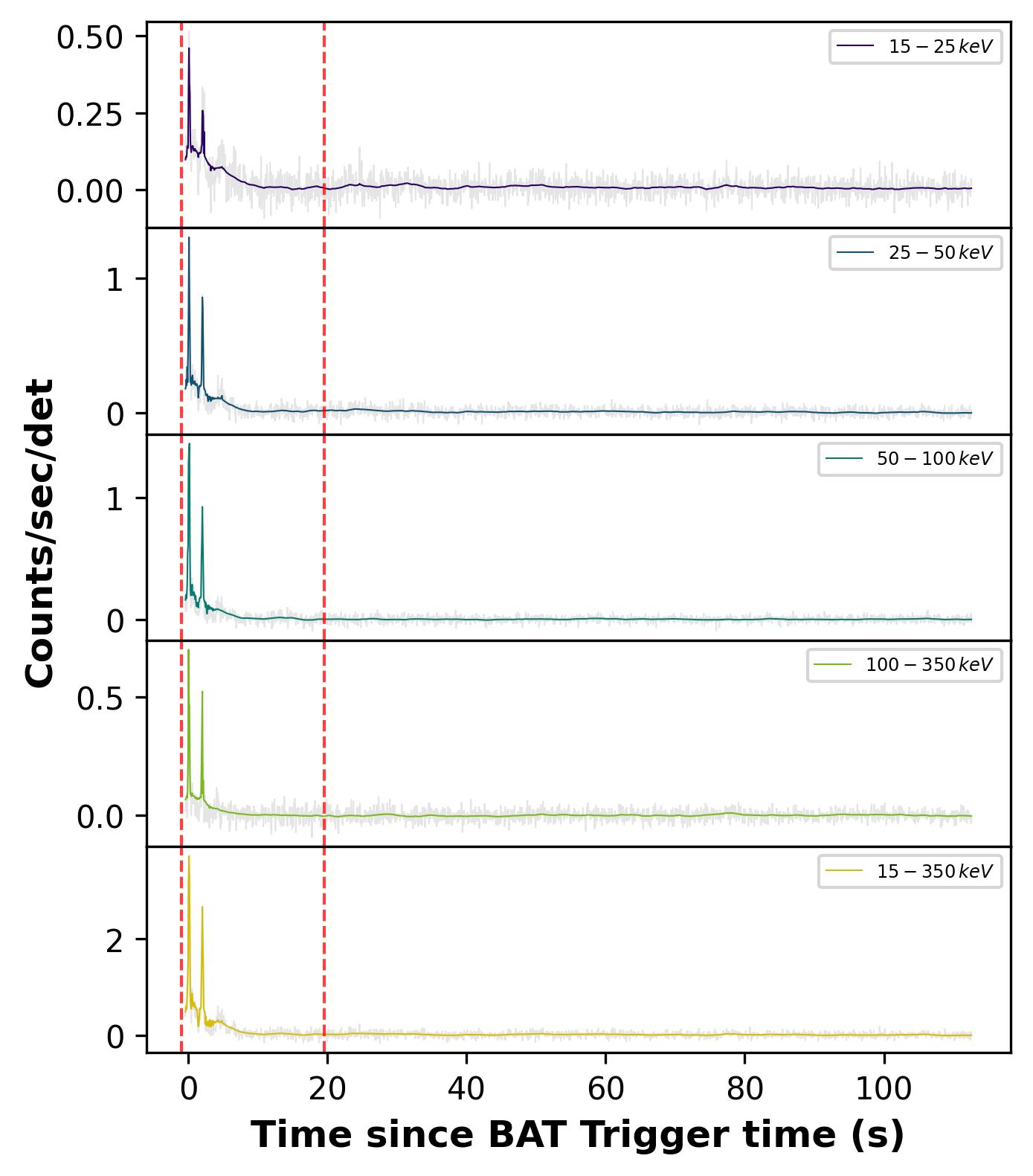}
     \label{fig:GRB 200716C}
     \caption{Light curves of GRB 200716C as extended emission GRB identified in this work. The light curves are taken from {\itshape Swift}/BAT (gray line) and its noise-reduced counterpart  (colored lines) were obtained with the {\sc FABADA} \citep{sanchez22} algorithm, limited to $T_{100}$. The extended emission region identified is illustrated with dashed red lines, with the lower temporal limit shifted $0.5$~seconds for better visualization of the main peak.}
 \end{figure}

\end{document}

%% file: table_classification.tex
\begin{table}[]
\begin{tabular}{p{2.5cm}p{1.2cm}p{1.2cm}p{1.2cm}}
\textbf{Name} & \textbf{This work 1} & \textbf{This work 2} & \textbf{Steinhardt} \\
\hline
GRB041217     & L                & L               & L                    \\
GRB041219C    & L                & L               & L                    \\
GRB041220     & L                & L               & L                    \\
GRB041223     & L                & L               & L                    \\
GRB041224     & L                & L               & L                    \\
GRB041226     & L                & L               & L                    \\
GRB041228     & L                & L               & L                    \\
GRB050117     & L                & L               & L                    \\
$\cdots$      &                  &                 &   \\             \hline
\end{tabular}
\caption{Classification of Swift Bursts obtained by t-SNE. ``L'' denotes LGRBs and ``S'' denotes SGRBs.  A full, machine-readable version is available online.}
\label{table:classification}
\end{table}

%% file: eeothers.tex
\begin{table*}
\centering
\begin{tabular}{p{3.0cm}p{6cm}rp{1.5cm}rp{1.5cm}}
\hline
\textbf{Reference} & \textbf{Criteria and comments}         & \textbf{GRBs} & \textbf{Instruments} & \textbf{Years}\\
\hline

\citet{kaneko15} & S/N ratio of binned LC&
  $\sim1700$ &
  BAT/GBM &
  $<2013$ \\
    \citet{lien16} &
  Qualitative: Used list from \citet{norris10} and eye inspection of GCN EE candidates  &
  $\sim 1000$ &
  BAT &
  $<2015$ \\

\citet{kisaka17}             & Phenomenological formulae: power-law decay fitting of Plateau and EE & $65$            & BAT/XRT             & $2005-2017$   \\

\citet{zhang20}             & S/N$>2$ in EE phase criterion      & $42\footnote{Sample of EE GRBs after S/N checking}$          &                  & $2005-2017$          \\

\citet{dichiara21}          & Signal adding  and S/N$\>1.5\sigma$ of binned LC following \citet{copete12}          & $8$             & BAT                 & $2005-2018$  \\

\citet{dainotti21}           & BAT searching  & $162$           & BAT                & $2005-2019$   &                                                      \\
\hline
\end{tabular}
\end{table*}

%% file: tableee2.tex
\begin{table}
\centering
\caption{GRBs reported with extended emission. References: (1) \cite{lien16}, (2) \cite{dichiara21}, (3) \cite{kaneko15}, (4) \cite{dainotti21}, (5) \cite{zhang20}, (6) \cite{kisaka17}, (7) \cite{Fong2022}}
\label{tab:eelist}
\begin{tabular}{lrrr}
\hline
\textbf{Name} & \textbf{T90 [s]} & \textbf{z} & \textbf{References}\\
\hline
GRB 050724       & 98.68                 & 0.26       & 1, 3, 4, 5, 6      \\
GRB 050911       & 16.24                 & 0.16*    & 4                  \\
GRB 051016B      & 4.00                  & 0.94       & 3                  \\
GRB 051016B      & 4.00                  & 0.94       & 4, 5               \\
GRB 051210       & 1.30                  & 2.58    & 2, 6,7               \\
GRB 051221A      & 1.39                  & 0.55       & 5, 6               \\
GRB 051227       & 115.40                & \nodata    & 1, 4, 5, 6         \\
GRB 060121       & \nodata               & \nodata    & 2                  \\
GRB 060306       & 60.94                 & 1.56       & 4, 5               \\
GRB 060313       & 0.74                  & $<$1.7    & 5, 6               \\
GRB 060607A      & 103.03                & 3.07       & 4                  \\
GRB 060614       & 109.10                & 0.13       & 3, 4, 5, 6         \\
GRB 060801       & 0.50                  & 1.13       & 5, 6               \\
GRB 060814       & 145.07                & 1.92       & 4, 5               \\
GRB 060912A      & 5.03                  & 0.94       & 4                  \\
GRB 061006       & 129.79                & 0.44       & 1, 3, 4, 5, 6      \\
GRB 061021       & 47.82                 & 0.35       & 4                  \\
GRB 061201       & 0.78                  & 0.41    & 5, 6, 7               \\
GRB 061210       & 85.23                 & 0.41       & 1, 3, 4, 5, 6      \\
GRB 070223       & 128.00                & \nodata    & 4, 5               \\
GRB 070506       & 5.99                  & 2.31       & 3, 4, 5            \\
GRB 070714B      & 65.64                 & 0.92       & 1, 3, 4, 5, 6      \\
GRB 070724A      & 0.43                  & 0.46       & 5, 6               \\
GRB 071227       & 142.48                & 0.38       & 1, 5, 6            \\
GRB 080123       & 114.91                & 0.50    & 5, 6, 7               \\
GRB 080503       & 176.00                & \nodata    & 1, 3, 6            \\
GRB 080603B      & 59.12                 & 2.69       & 4, 5               \\
GRB 080702A      & 0.51                  & \nodata    & 6                  \\
GRB 080905A      & 1.02                  & 0.12       & 5, 6               \\
GRB 080905B      & 120.94                & 2.37       & 5                  \\
GRB 080913       & 7.46                  & 6.73       & 4                  \\
GRB 080919       & 0.60                  & \nodata    & 6                  \\
GRB 081024A      & 1.82                  & \nodata    & 6                  \\
GRB 090426A       & 1.24                  & 2.61       & 5, 6               \\
GRB 090510       & 5.66                  & 0.90       & 5, 6               \\
GRB 090515       & 0.04                  & 0.40    & 6, 7                  \\
GRB 090531B      & 55.00                 & \nodata    & 1, 3               \\
GRB 090715A      & 64.00                 & \nodata    & 1                  \\
GRB 090916       & 62.45                 & \nodata    & 1                  \\
\hline
\end{tabular}
\end{table}

\begin{table}
\centering
\caption{Continuation: GRBs reported with extended emission. References: (1) \cite{lien16}, (2) \cite{dichiara21}, (3) \cite{kaneko15}, (4) \cite{dainotti21}, (5) \cite{zhang20}, (6) \cite{kisaka17}, (7) \cite{Fong2022}, (8) \cite{oconnor2022}, (9) \cite{Sakamoto2013}, (10) \cite{Yang2022}}
\label{tab:eelist1}
\begin{tabular}{lrrr}
\hline
\textbf{Name} & \textbf{T90 [s]} & \textbf{z} & \textbf{References}\\
\hline
GRB 090927       & 2.16                  & 1.37       & 3, 4, 5            \\
GRB 091109B      & 0.27                  & \nodata    & 6                  \\
GRB 100117A      & 0.29                  & 0.92       & 6                  \\
GRB 100212A      & 163.76                & \nodata    & 3                  \\
GRB 100522A      & 48.00                 & \nodata    & 3                  \\
GRB 100625A      & 0.33                  & 0.45       & 6                  \\
GRB 100702A      & 0.51                  & \nodata    & 6                  \\
GRB 100704A      & 196.88                & \nodata    & 4                  \\
GRB 100724A      & 1.39                  & 1.29       & 6                  \\
GRB 100814A      & 177.26                & 1.44       & 4                  \\
GRB 100816A      & 2.88                  & 0.80       & 4                  \\
GRB 100906A      & 114.63                & 1.73       & 4                  \\
GRB 101219A      & 0.83                  & 0.72       & 6                  \\
GRB 110207A      & 82.58                 & \nodata    & 3                  \\
GRB 110402A      & 56.21                 & 0.85    & 3, 8                  \\
GRB 111005A      & 23.21                 & 0.01    & 4                  \\
GRB 111117A      & 0.46                  & 2.21    & 6, 9                  \\
GRB 111121A      & 113.33                & \nodata    & 1, 3, 6            \\
GRB 111228A      & 101.24                & 0.72       & 4                  \\
GRB 120305A      & 0.10                  & 0.22    & 6, 7                  \\
GRB 120521A      & 0.51                  & \nodata    & 6                  \\
GRB 120804A      & 0.81                  & 1.30       & 2, 6               \\
GRB 121014A      & 80.00                 & \nodata    & 3                  \\
GRB 121226A      & 1.01                  & 1.37    & 6, 7                  \\
GRB 131004A      & 1.54                  & 0.72       & 6                  \\
GRB 140516A      & 0.26                  & 0.35    & 6, 7                  \\
GRB 140930B      & 0.84                  & 1.47   & 6, 7                  \\
GRB 150120A      & 1.20                  & 0.46       & 6                  \\
GRB 150301A      & 0.48                  & \nodata    & 6                  \\
GRB 150423A      & 0.22                  & \nodata    & 6                  \\
GRB 150424A      & 81.00                 & 0.30       & 1, 4, 6            \\
GRB 150831A      & 0.92                  & \nodata    & 6                  \\
GRB 151229A      & 1.44                  & 1.40   & 6, 8                  \\
GRB 160408A      & 0.32                  & 1.90    & 6, 7                  \\
GRB 160410A      & 96.00                 & 1.72       & 2, 4               \\
GRB 160525B      & 0.29                  & 0.64    & 6, 7                  \\
GRB 160624A      & 0.19                  & 0.48       & 6                  \\
GRB 160821B      & 0.48                  & 0.16       & 6                  \\
GRB 160927A      & 0.49                  & \nodata    & 6                  \\
GRB 161004A      & 1.32                  & \nodata    & 6                  \\
GRB 170127B      & 0.512                 & 2.28    & 6, 7                  \\
GRB 181123B      & 0.26                  & 1.75    & 2, 7                  \\
GRB 211211A      & 51.37                  & 0.08    & 10                  \\
\hline
\end{tabular}
\end{table}

%% file: candidates.tex
\begin{table}
\centering
\caption{GRBs identified as candidates to have extended emission. The information of duration $T_{90}$ and redshift $z$ were obtained from the {\itshape Swift}/BAT database.}
\label{tab:candidates}
\begin{tabular}{lrr}
\hline
\textbf{Name} & \textbf{$T_{90}$ [s]} & \textbf{z}\\
\hline
GRB 080123 & 114.9$\pm$55.3 & \nodata\\
GRB 140506A & 111.1$\pm$9.5 & 0.89 \\
GRB 170728B & 47.7$\pm$26.5 & \nodata \\
GRB 180618A & 47.4$\pm$11.2 & $<1.2$\\
GRB 180805B & 122.2$\pm$18.0 & \nodata \\
GRB 200219A & 288.0$\pm$50.6 & \nodata  \\
GRB 200716C & 86.6$\pm$15.2 &  \nodata\\
\hline
\end{tabular}
\end{table}